\documentclass[preprint,nofootinbib,aps,superscriptaddress,eqsecnum]{revtex4-1} 
 \pdfoutput=1
\usepackage{epsfig} 
\usepackage{amsmath} 
\usepackage{physics}
\usepackage{amssymb}
\usepackage{graphicx}
\usepackage{color}
\usepackage{float}
\usepackage{soul,xcolor}
\usepackage[font=small]{caption}
\usepackage{braket}
\usepackage{bbold}
\usepackage{subfigure}
\usepackage{braket}
\usepackage{titlesec}
\usepackage{placeins}
\usepackage{hyperref}
\usepackage{caption}
\usepackage{array}
\usepackage{booktabs}
\usepackage{subcaption}
\usepackage{changepage}
\usepackage{gensymb}
\usepackage{trimclip}
\usepackage{accents}
\usepackage{booktabs}
\usepackage{multirow}
\usepackage{lipsum}
\usepackage{placeins}
\usepackage{comment}

\PassOptionsToPackage{unicode}{hyperref}
\PassOptionsToPackage{naturalnames}{hyperref}

\usepackage[justification=centering]{caption}
\newcolumntype{P}[1]{>{\centering\arraybackslash}p{#1}}
\hypersetup{colorlinks,linkcolor={blue},citecolor={blue},urlcolor={blue}}

\def\bea{\begin{eqnarray}}
\def\eea{\end{eqnarray}}
\def\be{\begin{equation}}
\def\ee{\end{equation}}

\begin{document}

\title{The effect of non-standard interactions and environmental decoherence at DUNE}

\author{Chinmay Bera}
\email[Email Address: ]{chinmay20pphy014@mahindrauniversity.edu.in}
\affiliation{Department of Physics, \'Ecole Centrale School of Engineering - Mahindra University, Hyderabad, Telangana, 500043, India}

\author{K. N. Deepthi}
\email[Email Address: ]{nagadeepthi.kuchibhatla@mahindrauniversity.edu.in}
\affiliation{Department of Physics, \'Ecole Centrale School of Engineering - Mahindra University, Hyderabad, Telangana, 500043, India}

\author{Rukmani Mohanta}
\email[Email Address: ]{rmsp@uohyd.ac.in}
\affiliation{School of Physics, University of Hyderabad, Hyderabad - 500046, India}

\begin{abstract}
The Deep Underground Neutrino Experiment (DUNE) is a proposed long-baseline neutrino oscillation experiment that will project an on-axis wide-band neutrino beam over a distance of 1300 km to determine the unknowns in the neutrino sector. Given the baseline of 1300 km and the intense beam facility, DUNE is a promising experiment to study the sub-leading effects such as environmental decoherence, matter induced non-standard interactions (NSIs), neutrino decay, etc. In this study, we investigate how NSI and environmental decoherence affect the neutrino oscillation probabilities simultaneously. Considering the modified probabilities we obtain the updated mass hierarchy (MH) and CP violation (CPV) sensitivities of DUNE. 
Furthermore, we demonstrate the sensitivity of DUNE to distinguish between the effects of NSI and environmental decoherence.
\end{abstract}


\maketitle
\flushbottom
\section{Introduction}
Neutrino oscillation experiments are advancing into the era of precision measurements, paving the way to probe new physics phenomenon beyond the standard model. Non-standard interactions (NSIs)~\cite{Wolfenstein:1977ue,Guzzo:1991hi,Roulet:1991sm, Bergmann:1999rz, Hattori:2002uw, Blennow:2005qj}, environmental decoherence~\cite{Chang:1998ea,Benatti:2000ph,Gago:2000qc,Super-Kamiokande:2004orf,Farzan:2008zv}, neutrino decay~\cite{Bahcall:1972my,Schechter:1981cv,Gelmini:1983ea,Gonzalez-Garcia:2008mgl} and sterile neutrinos~\cite{Okada:1996kw,Conrad:2016sve,Barry:2011wb,Deepthi:2019ljo} are the prominent scenarios of new physics (NP) that are currently under investigation. In this paper, we report the effect of NSIs and environmental decoherence on the neutrino propagation through the Earth's crust at DUNE~\cite{DUNE:2016hlj}. 
The primary physics goals of this experiment~\cite{DUNE:2020jqi} include the precision measurement of $\delta_{CP}$ and the determination of mass hierarchy (MH), i.e, $\Delta m^2_{31} > 0$ (normal hierarchy-NH) and $\Delta m^2_{31} < 0$ (inverted hierarchy-IH). To accomplish these goals, DUNE considers a 1300 km baseline for a substantial matter effect, high intensity broad-band neutrino beam and better particle identification technology. This powerful instrumentation provides a significant opportunity to explore new physics.

In this work we study the impact of NSIs and decoherence on the neutrino propagation phenomenologically, without probing in detail the underlying causes of these NP scenarios. In addition to the standard matter effect, we consider that the propagation of neutrinos in matter is influenced by NSIs via neutral-current (NC) interactions~\cite{Meloni:2009ia,Liao:2016orc}. 
We neglect the source and detection NSIs (charge-current(CC) interaction)\footnote{Since model independent bounds on the charged current NSI involve charged leptons and these are an order of magnitude stronger than the neutral current NSI~\cite{Biggio:2009nt}.} in this study. 
Additionally, we assume that the neutrino system behaves as an open quantum system. In an open quantum system approach the coherent neutrino evolution can be influenced by the presence of environmental factors such as, the fluctuations of the space-time at Planck scale (quantum foam), quantum gravity, interaction with virtual black holes. This phenomenon is popularly known as environmental decoherence\footnote{Environmental decoherence is different from wave-packet decoherence~\cite{Giunti:1998kim, Blennow:2005ohl, Akhmedov:2012her, Chang:2016chu, Gouvea:2021rom}.}. This effect alters the oscillation probabilities by introducing a damping term $e^{-\Gamma L}$~\cite{Coelho:2017byq,Bera:2024hhr}, where $\Gamma$ is the decoherence parameter and $L$ is the baseline.
Based on the different origins of the decoherence phenomenon the $\Gamma$ parameter is subject to different power-law dependencies~\cite{Romeri:2023cgt} of neutrino energy $E$ i.e. $\Gamma \propto E^n$, where $n$ is the power-law index. This suggests that an experiment having relatively longer baseline, considerable matter effect and high energy neutrino beam is suitable to study the propagation NSIs and environmental decoherence.

The individual impact of these two new physics scenarios have been studied extensively in the context of solar neutrinos, atmospheric neutrinos, accelerator neutrinos and supernova neutrinos. Here, we provide several relevant references in the context of DUNE. NSIs at DUNE have been investigated in refs.~\cite{deGouvea:2015ndi,Coloma:2015kiu,Masud:2016bvp,C:2016nrg,Masud:2016nuj,Blennow:2016etl,Agarwalla:2016fkh,Deepthi:2016erc,Deepthi:2017gxg,Meloni:2018xnk,Verma:2018gwi,DUNE:2020fgq,Bakhti:2020fde,Chatterjee:2021wac,Brahma:2023wlf}, whereas in refs.~\cite{Carrasco:2018sca,Carpio:2019mass,Gomes:2019for,Romeri:2023cgt} the authors have illustrated the effect of environmental decoherence on neutrino oscillations. 
In this study, we present a novel analysis of the combined impact of NSIs and environmental decoherence on the MH and CPV sensitivity of DUNE for the first time. Moreover, if the simulated data is generated assuming more than one new physics scenario then it is crucial to verify whether they can be differentiated from each other or not~\footnote{In refs.~\cite{Denton:2022pxt,Giarnetti:2024mdt,Calatayud-Cadenillas:2024wdw} authors have explored the sensitivity of DUNE to differentiate between NP scenarios.}. In this context, to distinguish between NSI and environmental decoherence using DUNE, we incorporate one NP scenario in the simulated data and test over the other.

This paper is organized as follows. In section~\ref{sec:math-form}, we provide a brief overview of the derivation of oscillation probabilities in the presence of NSIs and decoherence. In section~\ref{sec:expt-detail} and section~\ref{sec:stat-ana}, we describe simulation details and statistical analysis respectively. In section~\ref{sec:results}, we showcase and discuss our results. We summarize and conclude our findings in section~\ref{sec:conclusion}.

\section{Mathematical formulation}\label{sec:math-form}

The effective Lagrangian in the presence of neutral current (NC) NSI can be described by the effective four-fermion operators~\cite{Wolfenstein:1977ue,Guzzo:1991hi}
\begin{equation}\label{eq:lagrangian-nsi}
    \mathcal{L}_{NC-NSI} = -2\sqrt{2}G_F\epsilon_{\alpha\beta}^{fC}\left(\bar{\nu}_\alpha \gamma^\mu P_L \nu_{\beta}\right) \left(\bar{f}\gamma_\mu P_C f\right),
\end{equation}
here, $G_F$ is the Fermi constant and $\epsilon_{\alpha\beta}^{fC}$ are the NSI parameters, with $\alpha, \beta = e, \mu, \tau$; $f = e, u, d$ (first generation fermion); $P_L$ and $P_R$ denote the left and right chiral projection operators respectively where $C = L, R$. Using NSI operators of dimension-six, the non-standard parameters (estimate the deviation from standard model) are related to the mediator mass ($m_{NSI}$) as $\epsilon \sim (g_{NSI}^2/m_{NSI}^2)G_F^{-1}$, where $g_{NSI}$ is the coupling of the new interaction. 
We define the NSI parameters as
$\epsilon_{\alpha\beta} = \sum_{f,C} \frac{n_f}{n_e}\epsilon_{\alpha\beta}^C~$,
where, $n_f$ represents the number density of fermions (flavor $f$) and $n_e$ is the number density of electrons.

The effective Hamiltonian of the system in the flavor basis, in the presence of standard matter effect ($H_{matter}$) and NSIs ($H_{NSI}$) is 
\begin{equation}\label{eq:H_eff}
\begin{aligned}
     H_{eff} &= H_0 + H_{matter} + H_{NSI}\\
     &= \frac{1}{2E}\left[U\begin{pmatrix}
         0 & 0 & 0 \\
         0 & \Delta m^2_{21} & 0 \\
         0 & 0 & \Delta m^2_{31}
     \end{pmatrix}U^\dagger + \begin{pmatrix}
         A & 0 & 0 \\
         0 & 0 & 0 \\
         0 & 0 & 0 
     \end{pmatrix}  + A \begin{pmatrix}
         \epsilon_{ee} & \epsilon_{e\mu} & \epsilon_{e\tau} \\
         \epsilon_{\mu e} & \epsilon_{\mu\mu} & \epsilon_{\mu\tau} \\
         \epsilon_{\tau e} & \epsilon_{\tau\mu} & \epsilon_{\tau\tau}
     \end{pmatrix} \right]~,
\end{aligned}
\end{equation}
where $U$ is the standard PMNS matrix, $A=2\sqrt{2}G_Fn_e E$ and $\epsilon_{\beta\alpha} = \epsilon_{\alpha\beta}^* = |\epsilon_{\alpha\beta}|e^{-i\delta_{\alpha\beta}}$. We mainly focus on the two off-diagonal NSI parameters, $\epsilon_{e\mu}$ ($= |\epsilon_{e\mu}|e^{i\delta_{e\mu}}$) and $\epsilon_{e\tau}$ ($= |\epsilon_{e\tau}|e^{i\delta_{e\tau}}$) as they are less constrained and assume all other parameters to be zero.

The diagonalized Hamiltonian in the mass basis 
\begin{equation}
    H_{eff}^m = \frac{1}{2E}\left(0,\Delta \tilde{m}^2_{21},\Delta \tilde{m}^2_{31}\right)~,
\end{equation}
here $\Delta \tilde{m}^2_{jk}$ ($j,k = 1,2,3$) are the modified mass-squared differences, and the modified PMNS matrix $\tilde{U}$ is responsible for conversion between flavor and mass basis. In general, the NSI parameters ($|\epsilon_{\alpha \beta}|e^{i\delta_{\alpha\beta}}$, $\alpha,\beta = e, \mu, \tau$) appear in modified mixing matrix $\tilde{U}$ and $\Delta \tilde{m}_{jk}^2$ (see ref.~\cite{Meloni:2009ia}). However, one can note from eq.~(\ref{eq:Delta_m_tilda}) of Appendix~\ref{app_mixing_mass_square}  that $\Delta \tilde{m}_{jk}^2$ is free from NSI parameters in our case, where we only considered $\epsilon_{e\mu}$ and $\epsilon_{e\tau}$ to be non-zero. The relevant analytical forms of $\tilde{U}$ and $\Delta \tilde{m}_{jk}^2$ using first order perturbation theory are provided in Appendix~\ref{app_mixing_mass_square} for analysis purposes.

Upon obtaining the Hamiltonian of the neutrino system with matter NSI, we proceed further to incorporate decoherence, considering 
an open quantum system scenario. Neutrino propagation in the framework of open quantum system undergoes a loss of quantum coherence due to the interaction with the stochastic environment. This loss is introduced using Lindblad master equation which represents the time evolution of the density matrix as given below~\cite{Lindblad:1976g,GKS:1976vit}

    \begin{equation}
    \frac{d\rho^m(t)}{dt} = -i\left [H_{eff}^m,\rho^m(t) \right] + \mathcal{D} \left[ \rho^m (t) \right],
    \label{eq:LME}
\end{equation}
here $\rho^m$ is the density matrix of the neutrino mass eigen states and $H_{eff}^m$ is the respective Hamiltonian. The dissipative matrix  $\mathcal{D}[\rho^m (t)]$ represents the interaction between the neutrino system and the environment. We parameterize $\mathcal{D}[\rho^m (t)]$ by imposing some generic properties of the density matrix such as complete positivity and preserving trace normalization. Complete positivity condition gives the Lindblad form of the dissipator as~\cite{GKS:1978g} 
\begin{equation}
\begin{aligned}
    \mathcal{D} \left[ \rho^m (t) \right] &= \frac{1}{2}\sum_{n = 1}^{N^2 - 1} \left\{[\mathcal{V}_n , \rho^m \mathcal{V}_n^\dagger] + [\mathcal{V}_n \rho^m , \mathcal{V}_n^\dagger]\right\},
    \end{aligned}
        \label{eq:D-term}
\end{equation}
\noindent

where $N$ is the dimension of the Hilbert space and $\mathcal{V}_n$ are the interaction operators. Further, we impose an increase in von Neumann entropy $ S = - Tr(\rho^m \ln \rho^m) $~\cite{Banks:1984bsp,Benatti:1988nar} and the conservation of average energy $Tr(\rho^m H_{eff}^m)$ of the neutrino system and obtain the time evoluted density matrix. The diagonal ($\rho^m_{jk}, j=k$) and off-diagonal ($\rho^m_{jk}, j \neq k$) matrix elements of the density matrix are given by
\begin{equation}
\begin{aligned}
    & \rho^m_{jj}(t) = \rho_{jj}(0),\\
    & \rho^m_{jk}(t) = \rho_{jk}(0) \exp-(\Gamma_{jk} + i \Delta_{jk}).
    \end{aligned}
\end{equation}
with 
\begin{equation}
    \Gamma_{jk} = \Gamma_{kj} = \frac{1}{2}\sum_{n = 1}^8 (d_{n,j} - d_{n,k})^2
    \label{eq:Gamma_ij},
\end{equation}
where, $d_{n,j}$, $d_{n,k}$ are the diagonal elements of $\mathcal{V}_n$ operator and $j,k$ take the values $1,2,3$. We refer \cite{Gomes:2019for} for comprehensive details.

We consider the modified-mixing matrix ($\tilde{U}$) to convert the density matrix from mass basis to flavor basis using $\tilde{\rho}^{\alpha} = \tilde{U}~\tilde{\rho}^m~\tilde{U}^\dagger$. The neutrino transition probability from initial flavor '$\nu_\alpha$' to final flavor '$\nu_\beta$' in terms of density matrix is obtained using
\begin{equation}
   \begin{aligned}
        P_{\alpha \beta}(t) &= Tr[\tilde{\rho}^\alpha (t) \tilde{\rho}^\beta (0)].
   \end{aligned}
   \label{eq:P1}
\end{equation}
The explicit form of the transition probability assuming ultra-relativistic neutrinos ($t \approx L$) is given by~\cite{Gomes:2019for,Coloma:2018ice}
\begin{equation}
\begin{aligned}
        P_{\alpha \beta}(L) &= \delta_{\alpha \beta} - 2\sum_{j > k} Re \left( \tilde{U}_{\beta j} \tilde{U}_{\alpha j}^* \tilde{U}_{\alpha k} \tilde{U}_{\beta k}^* \right) + 2\sum_{j > k} Re \left( \tilde{U}_{\beta j} \tilde{U}_{\alpha j}^* \tilde{U}_{\alpha k} \tilde{U}_{\beta k}^* \right) \exp(-\Gamma_{jk} L) \cos(\frac{\Delta \tilde{m}_{jk}^2}{2E}L) \\& + 2\sum_{j > k} Im \left( \tilde{U}_{\beta j} \tilde{U}_{\alpha j}^* \tilde{U}_{\alpha k} \tilde{U}_{\beta k}^* \right) \exp(-\Gamma_{jk} L) \sin(\frac{\Delta \tilde{m}_{jk}^2}{2E}L)~.
        \end{aligned}
        \label{eq:Pab}
\end{equation}
 
In eq.~(\ref{eq:Pab}),  $\Gamma_{jk}$ appears in the decoherence term $e^{-\Gamma_{jk}L}$, which damps the oscillation probability. Substituting $\Gamma_{jk} = 0$ into the eq.~(\ref{eq:Pab}) results in the oscillation probabilities with matter NSIs. Alternatively, considering  $\epsilon_{\alpha\beta} = 0$ in eq.~(\ref{eq:Pab}) gives the oscillation probabilities for decoherence. Including non-zero $\epsilon_{\alpha\beta}$ and $\Gamma_{jk}$ together helps us to explore the combined effect of NSI and decoherence on the neutrino oscillations. Throughout our analysis we consider $\Gamma_{21} = \Gamma_{31} = \Gamma_{32} = \Gamma$.

We consider a general power law dependency of $\Gamma$ on the neutrino beam energy given by 
\begin{equation}
    \Gamma (E_\nu) = \Gamma_0 \left(\frac{E_\nu}{E_0}\right)^n~,
    \label{G-powerlaw}
\end{equation}
where, $\Gamma_0$ is a constant, $E_0$ is the reference energy taken as 1 GeV and $n = 0,\pm 1,\pm 2$. 
From eq.~(\ref{G-powerlaw}) one can note that for $E_\nu > E_0$, decoherence plays significant role when $n \geq 0$.

\section{Experimental simulation details}\label{sec:expt-detail}

The upcoming DUNE experiment is being installed in the USA and is expected to start collecting data in 2027. The experiment is composed of two detectors. Near detector complex~\cite{DUNE:2021tad} is located at the Fermilab site while a LAr-TPC (liquid argon time projection chamber) far detector (FD)~\cite{DUNE:2020ypp} of volume 40 kt is placed at Sanford Underground Research Facility which is $\sim$ 1285 km far from the neutrino source. The main physics goals of this experiment is to maximize the sensitivity in measuring the CP-violation (CPV) phase, determining the mass ordering and identifying the octant of $\theta_{23}$. To achieve this sensitivity, an on-axis neutrino beam is generated from 120 GeV protons with a beam power 1.2 MW, and projected to the far detector. The resulting flux is a broad-band beam (mostly $1-5 $ GeV) with a peak $\sim$ 2.5 GeV which corresponds to first oscillation maximum in the atmospheric oscillations. These neutrinos are detected at the far detector after traveling through the Earth's crust with average matter density of 2.848 $g/cm^3$. Changing the polarity of the horn current, DUNE can run in both neutrino and anti-neutrino modes. 

To obtain simulated data we use ``General Long Baseline Experiment Simulator (GLoBES)" software~\cite{Huber:2004ka, Huber:2007ji}, and the auxiliary files from ref.~\cite{DUNE:2021cuw}. We assume a total run time of 10 years which consists of 5 years in neutrino mode and 5 years in anti-neutrino mode ($5(\nu) + 5(\bar{\nu})$). This corresponds to $1.1 \times 10^{21}$ POT/year and a net exposure of 480 kton-MW-years. 

We define a new probability engine in GLoBES to implement the combined effect of NSI and decoherence. For the numerical diagonalization of the Hamiltonian we employ Cardano's $3 \times 3$ eigen-problem algorithm as described in ref.~\cite{Kopp:2006wp}.

\begin{table}[!htbp]\centering
\begin{tabular}{ |c|c|c| } 
 \hline\hline
 Parameters & True values & $3\sigma$ ranges \\[0.5ex]
 \hline
 $\sin^2{\theta_{12}}$ & $0.307$ & Fixed \\ [0.5ex]
 $\sin^2{\theta_{13}}$ & $0.02203$ & Fixed \\ [0.5ex]
 $\sin^2{\theta_{23}}$ & $0.572$ & $[0.407 : 0.623]$ \\ [0.5ex]
 $\delta_{CP}$ & $198\degree$ & $[0 : 360\degree]$ \\ [0.5ex]
 $\frac{\Delta m^2_{21}}{10^{-5}~eV^2}$ & $7.41$ & Fixed \\ [0.5ex]
 $\frac{\Delta m^2_{31}}{10^{-3}~eV^2}$ (NH) & $2.512$ & $[2.428 : 2.597]$ \\ [0.5ex]
 $\frac{\Delta m^2_{31}}{10^{-3}~eV^2}$ (IH) & $-2.498$ & $[-2.584 : -2.413]$\\[0.5ex]
 $\epsilon_{e\mu}$ & $0.1$ & $[-0.3:0.3]$~\cite{Biggio:2009nt} \\[0.5ex]
 $\epsilon_{e\tau}$ & $0.1$ & $[-0.19:0.13]$~\cite{Super-Kamiokande:2011dam} \\[0.5ex]
 $\Gamma_0$ [GeV] & $10^{-23}$ & $[10^{-27} : 10^{-22}]$~\cite{dosSantos:2023skk,DeRomeri:2023dht} \\[0.5ex]
 \hline\hline
\end{tabular}
 \caption{Standard oscillation parameters and $3\sigma$ ranges have been considered in our analysis are taken from NuFIT 5.3 (2024)~\cite{Esteban:2020cvm}. The representative values of non-standard parameters and their test ranges are also tabulated in the last three rows.}
 \label{table:parameters}
\end{table}

\section{Statistical analysis}\label{sec:stat-ana}
For the statistical analysis we consider the $\nu_\mu (\bar{\nu}_\mu) \rightarrow \nu_e(\bar{\nu}_e)$ appearance and $\nu_\mu(\bar{\nu}_\mu) \rightarrow \nu_\mu(\bar{\nu}_\mu)$  disappearance channels as these are relevant channels to demonstrate MH and CPV sensitivity. 
We refer to~\cite{DUNE:2021cuw,DUNE:2020jqi} for efficiencies and background specifications.

We study CPV and MH sensitivity in the presence of environmental decoherence, NSI, and combined effect of both NSI and decoherence. In addition, we differentiate between NSI and decoherence taking one in true and the other in test hypotheses. We obtain the results using Poissonian $\chi^2$ function which reads as

\begin{equation}
    \chi^2=\min _{\alpha_s,\alpha_b} \sum_{\text {channels }} 2 \sum_i\left[N_i^{\mathrm{test}} - N_{i}^{\mathrm{true}}+N_{i}^{\mathrm{true}} \log \left(\frac{N_{i}^{\mathrm{true}}}{N_{i}^{\mathrm{test}}}\right)\right] + \alpha_s^2 + \alpha_b^2~,
    \label{eq:chi-sq}
\end{equation} 
where, $N_i^{\mathrm{true}}$ and $N_i^{\mathrm{test}}$ are the number of true and test events (signal + background) in $i$-th bin, respectively. $\alpha_s$ and $\alpha_b$ are the pull terms to introduce signal and background normalization errors respectively~\cite{Fogli:2002pt, Huber:2002mx}.


\section{Results and discussion}\label{sec:results}\label{sec:results}
In this section, we present the results and the corresponding implications of our analysis, assuming the presence of non-zero $\epsilon_{e \mu}$, $\epsilon_{e \tau}$ and decoherence parameter $\Gamma$ ($\Gamma_{21} = \Gamma_{31} = \Gamma_{32}$). Additionally, we adopt energy power-law dependencies on $\Gamma$ as in eq.~(\ref{G-powerlaw}). Further, among the $\epsilon_{e\mu}$ and $\epsilon_{e\tau}$, we consider only one non-zero off-diagonal parameter at a time and neglect all other NSI parameters in eq.~(\ref{eq:H_eff}).  
We consider the standard and the non-standard parameter values provided in table~\ref{table:parameters}. Here, we would like to emphasize that all the results presented in section~\ref{sec:results} have been obtained numerically (without approximation).

\begin{table}[!htbp]\centering
\begin{tabular}{ |c|c|c|c|c|c|c| } 
 \hline\hline
 Channels & SM & $\epsilon_{e\tau}$ & $\epsilon_{e\mu}$ & $\Gamma$ & $\epsilon_{e\tau} + \Gamma$ & $\epsilon_{e\mu}+\Gamma$ \\[0.5ex]
 \hline
 $\nu_e$-appearance & $1762$ & $1356$ & $1565$ & $1809$ & $1368$ & $1627$ \\ [0.5ex]
 $\bar{\nu}_e$-appearance & $271.76$ & $307.7$ & $408.6$ & $264.85$ & $302.63$ & $400.47$ \\ [0.5ex]
 $\nu_\mu$-disappearance & 14274 & 14342 & 14384 & 14507 & 14617 & 14608 \\[0.5ex]
 $\bar{\nu}_{\mu}$-disappearance & $4502.28$ & $4483.2$ & $4435.14$ & $4665.53$ & $4648$ & $4596.86$ \\[0.5ex]
 \hline\hline
\end{tabular}
 \caption{Number of appearance and disappearance events considering NH. We assume no energy dependency on the decoherence parameters, i.e., $n=0$.}
 \label{table:event}
\end{table}

\subsection{Impact of $\epsilon_{e\tau}$ and $\Gamma$ at DUNE far detector}
In this subsection, we illustrate the modification of the oscillation probabilities assuming $\epsilon_{e\tau}$ and $\Gamma$ to be nonzero. Considering these parameters in the true and fit hypothesis we show the change in MH and CPV sensitivities of DUNE.
\subsubsection{Oscillation probabilities and events rate}

Oscillation probabilities could reveal information about new interactions of neutrinos with unknown fields.  
Due to these interactions oscillation probabilities can be rewritten in terms of new sets of coupling parameters. Eq.~(\ref{eq:Pab}) represents the transition probabilities in the presence of NSI and decoherence. First, we assume only $\epsilon_{e\tau} = |\epsilon_{e\tau}|e^{i\delta_{e\tau}}$ to be non-zero and $\Gamma_{21} = \Gamma_{31} = \Gamma_{32}~(= \Gamma)$ with no energy dependency ($n = 0$ in eq.~\ref{G-powerlaw}). To compute the probabilities and events rate we consider normal hierarchy (NH) and the representative values of nonstandard parameters as $|\epsilon_{e\tau}| = 0.1, \delta_{e\tau} = 0$ and $\Gamma = 10^{-23}$ GeV.

\begin{figure*}[htb!]
\includegraphics[width=0.485\linewidth]{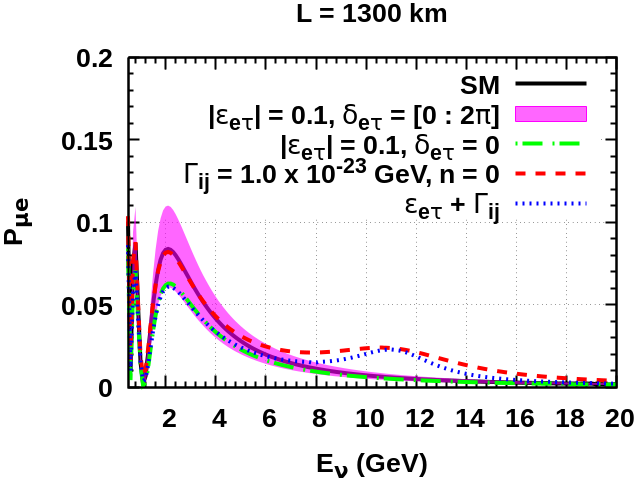}
\includegraphics[width=0.485\linewidth]{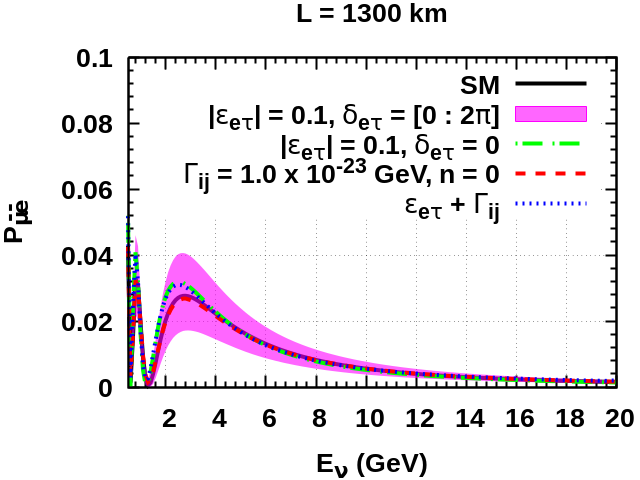}
\includegraphics[width=0.485\linewidth]{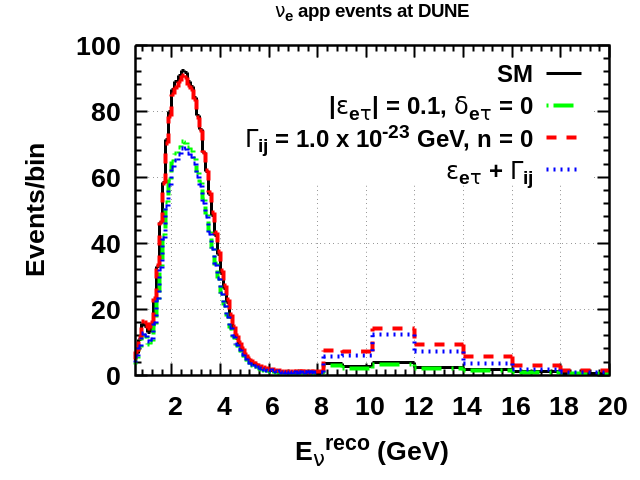}
\includegraphics[width=0.485\linewidth]{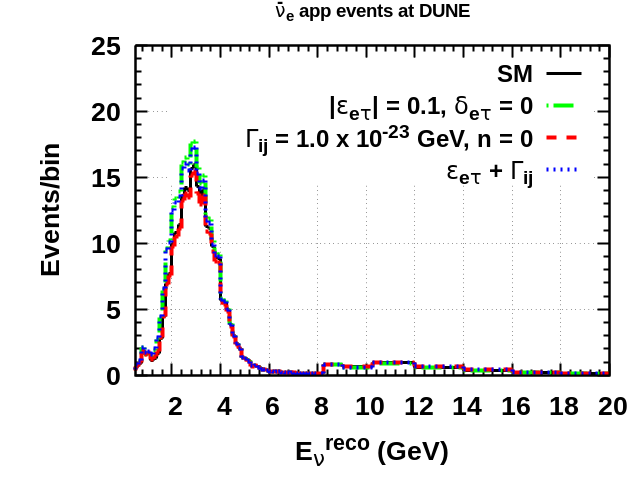}
\caption{Appearance probability (upper row) and appearance events spectrum (lower row) in the presence of $\epsilon_{e\tau}$ and $\Gamma$. Left and right panels correspond to neutrinos and antineutrinos respectively.}
\label{fig:prob-event-app-eps_etau}       
\end{figure*}

\noindent{\bf Appearance probability:}
In the upper row of fig.~\ref{fig:prob-event-app-eps_etau} we present the impact of $\epsilon_{e\tau}$ and $\Gamma$ on the $\nu_e$-appearance (left) and $\bar{\nu}_e$-appearance (right) channels. The curves in each plot represent the probabilities corresponding to the standard three flavor oscillations (black solid), NSI with $|\epsilon_{e\tau}| = 0.1$ and $\delta_{e\tau} = [0,2 \pi]$ (magenta band), NSI with $|\epsilon_{e\tau}| = 0.1$, $\delta_{e\tau} = 0$ (green dot-dashed), only decoherence ($\Gamma = 10^{-23}$ GeV)(red dashed) and the combined effect (blue dotted). As we can see from top left panel $\epsilon_{e\tau}$ modifies the $P_{\mu e}$ at energy $\sim~3$ GeV and the non-zero decoherence causes a peak\footnote{Please see ref.~\cite{Bera:2024hhr} for detail description of the peak $\sim 11$ GeV in the presence of decoherence.} at comparatively higher energy $\sim~11$ GeV. In the combined scenarios of both NSI and decoherence, the probability (blue dashed curve) is driven by green curve at lower energy and red curve at higher energy. On the other hand $P_{\bar{\mu} \bar{e}}$ channel shows a slight increase in the probability around $E_{\nu} \sim 2.5$ GeV in the presence of $\epsilon_{e\tau}$ and a negligible change in the presence of decoherence. As a result the combined effect leads to a small increase in $P_{\bar{\mu}\bar{e}}$ around the energy 2.5 GeV.

\noindent{\bf Appearance event:} 
In the lower row of fig.~\ref{fig:prob-event-app-eps_etau} we show the appearance events. The left and right plots represent $\nu_e$-appearance and $\bar{\nu}_e$-appearance events per bin. Each plot shows black, green, red and blue curves that represent the events rate considering standard interaction (SM), only NSI ($|\epsilon_{e\tau}| = 0.1, \delta_{e\tau} = 0$), only decoherence ($\Gamma = 10^{-23}$ GeV with $n = 0$) and combined effect ($\epsilon_{e\tau} + \Gamma$), respectively. The $\nu_e$ and $\bar{\nu}_e$-appearance events rate follow the similar trend as the probability. We tabulate total number of events from each channel in table~\ref{table:event}.

\begin{figure*}[t!]
\includegraphics[width=0.485\linewidth]{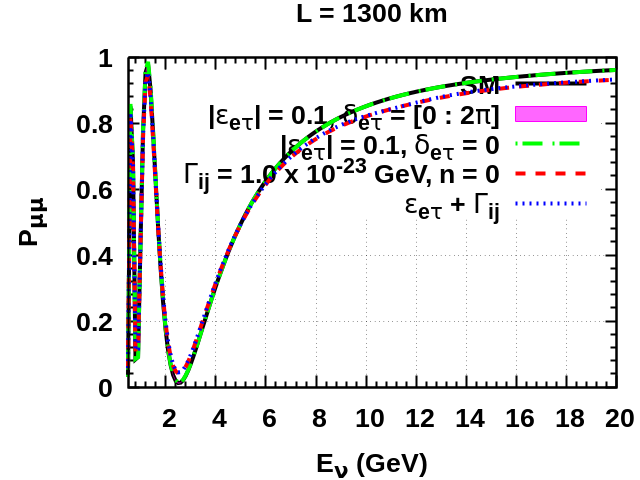}
\includegraphics[width=0.485\linewidth]{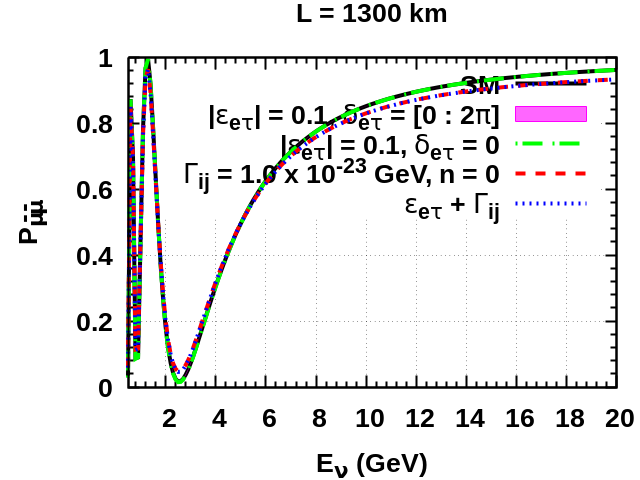}
\includegraphics[width=0.485\linewidth]{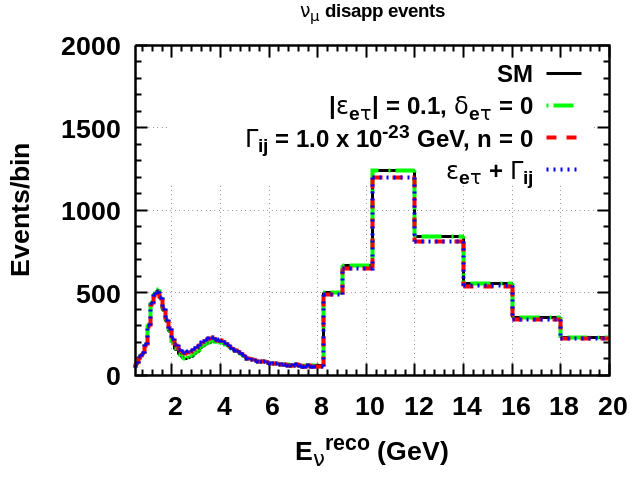}
\includegraphics[width=0.485\linewidth]{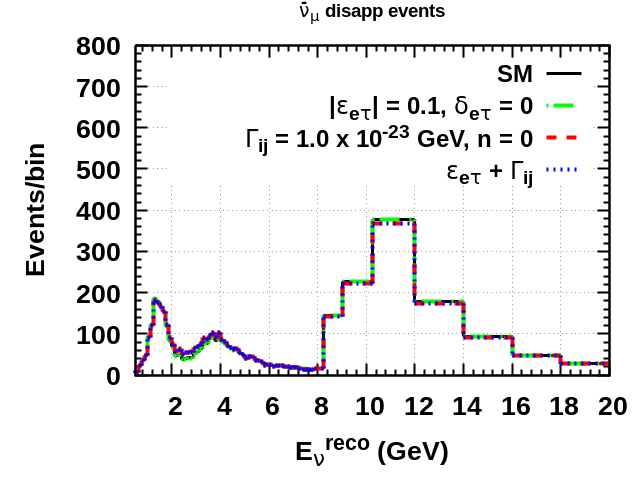}
\caption{Disappearance probability (upper row) and disappearance events spectrum (lower row) in the presence of $\epsilon_{e\tau}$ and $\Gamma$. Left and right panels correspond to neutrinos and antineutrinos respectively.}
\label{fig:prob-event-disapp-eps_etau}       
\end{figure*}

\noindent{\bf Disappearance probability:} In fig.~\ref{fig:prob-event-disapp-eps_etau} we present the disappearance channels and corresponding events rate. The left and right plot in the upper row represent the probabilities for $\nu_\mu \rightarrow \nu_\mu$ and $\bar{\nu}_\mu \rightarrow \bar{\nu}_\mu$ channels. In each plot we display four different cases as mentioned in the legends. As in ref.~\cite{Kopp:2007ne,Liao:2016orc}, we see that disappearance probabilities are independent of the $\epsilon_{e\tau}$, therefore SM curve (black) and the curve with $\epsilon_{e\tau}$ (green) show identical behaviour. There is small variation of the probability w.r.t to SM around $\sim~2.5$ GeV and at high energy $>~8$ GeV in the presence of $\Gamma$ and $n = 0$ (red dashed). 
The red dashed curve (only decoherence) and the blue dotted curve (NSI+decoherence) overlap with each other, indicating that the effect of $\epsilon_{e\tau}$ on the disappearance channels $\nu_\mu$ and $\bar{\nu}_\mu$ is almost negligible. The disappearance events spectra also follow the similar trend as the corresponding probabilities.

\FloatBarrier
\subsubsection{Mass hierarchy and CP-violation sensitivities}

    \begin{figure*}
    \centering
    \includegraphics[width=0.485\linewidth]{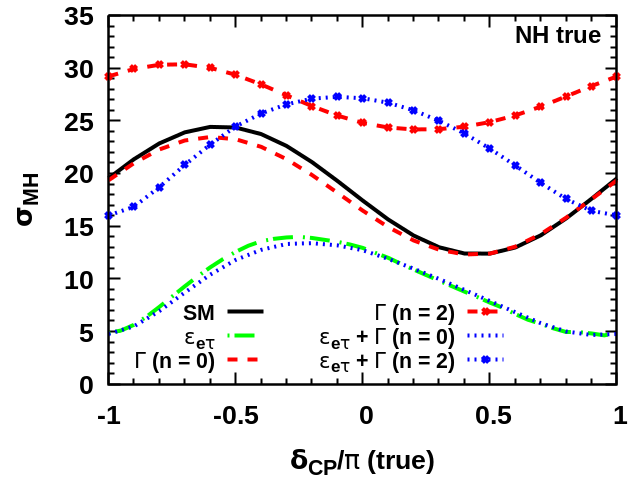}
    \includegraphics[width=0.485\linewidth]{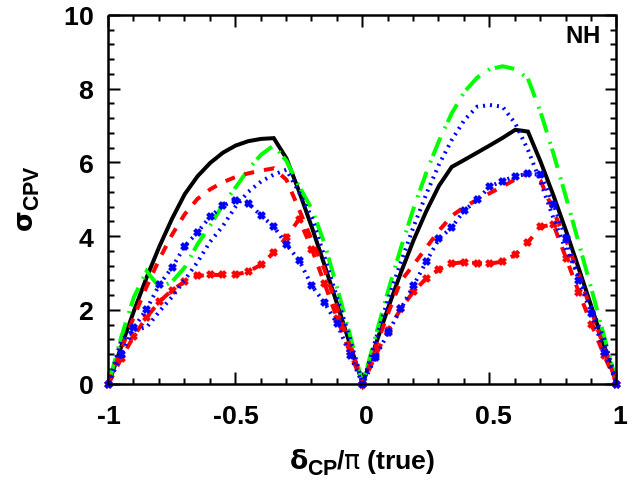}
    \includegraphics[width=0.485\linewidth]{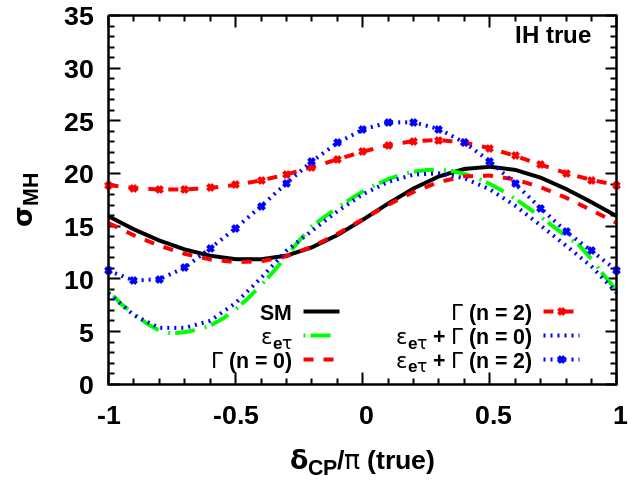}
    \includegraphics[width=0.485\linewidth]{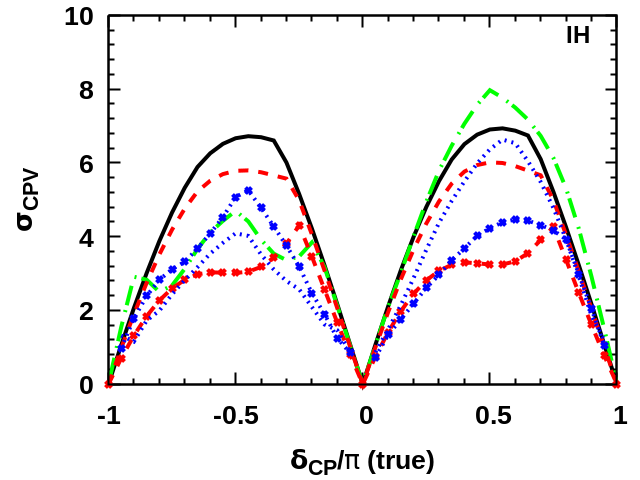}
    \caption{MH and CPV sensitivity vs $\delta_{CP}$. Upper and lower panels assume the true hierarchy to be NH and IH respectively. In both the panels we consider true value of $\epsilon_{e\tau} = 0.1$, $\Gamma_0 = 10^{-23}$ GeV.}
    \label{fig:cpv-mh-sensitivity-eps_etau}
\end{figure*}

Two major scientific objectives of DUNE are to determine mass hierarchy and estimate $\delta_{CP}$ to the utmost precision. In this context, we analyze both these sensitivities in the presence of non-zero NSI, decoherence, and the combined phenomenon.

During propagation from source to detector through Earth's crust, a matter potential ($A$) arises due to the forward scattering of electron neutrinos ($\nu_e$) with electrons in the Earth matter. The matter effect causes a resonance ($\Delta m^2_{31}/2E \approx A$) and enhances the probabilities of neutrino or antineutrino oscillations depending on the hierarchy. In NH (IH), resonance occurs for neutrinos (antineutrinos).
NSIs modify the effective matter potential and reform the oscillation probabilities to lower or higher than the probability with standard matter effect.
This leads to ambiguity in identifying the true hierarchy. In addition, the degeneracies between the new physics parameters and the standard parameters cause a difficulty in determining MH.

In the left panel of fig.~\ref{fig:cpv-mh-sensitivity-eps_etau}, we present MH sensitivity. Upper and lower panels show the results assuming true NH and IH respectively. In each plot we show the curves corresponding to SM (black), individual effects of $\epsilon_{e\tau}$ (green), $\Gamma$ (red), and their combined effect (blue). The red and blue line point curves represent the energy dependency on $\Gamma$ with power-law index $n = 2$ for decoherence in the individual and combined effect. We define MH sensitivity as
\begin{equation}\label{eq:MH}
\begin{aligned}
    \Delta \chi_{MH}^2 = \chi^2_{true}(\Delta m^2_{31} > 0, \epsilon^{true}_{e\tau}, \Gamma^{true}) - \chi^2_{test}(\Delta m^2_{31} < 0, \epsilon^{test}_{e\tau}, \Gamma^{test})~{\text{for NH true}},\\
    \Delta \chi_{MH}^2 = \chi^2_{true}(\Delta m^2_{31} < 0, \epsilon^{true}_{e\tau}, \Gamma^{true}) - \chi^2_{test}(\Delta m^2_{31} > 0, \epsilon^{test}_{e\tau}, \Gamma^{test})~{\text{for IH true}},
    \end{aligned}
\end{equation}
and marginalize over $\theta_{23}$, $\delta_{CP}$, $\Gamma$, $\epsilon_{e\tau}$. From upper left plot we see that there is no significant change in MH sensitivity (with respect to SM) in the presence of decoherence with $n = 0$, while in the presence of NSI ($\epsilon_{e\tau}$) we observe a considerable reduction in MH sensitivity compared to the SM case. Therefore, the combined effect ($\epsilon_{e\tau} + \Gamma$ ($n = 0$)) significantly reduces the sensitivity to MH. On the other hand the $\Gamma$ with $n = 2$ increases the sensitivity. Hence, the combined effect with $\epsilon_{e\tau} + \Gamma~(n = 2)$ improves the sensitivity compared to the case $\epsilon_{e\tau} + \Gamma~(n = 0)$.

For the case of true IH and test NH (lower left) we observe that the effect of decoherence in the $n = 0$ case (red dashed) is not noteworthy, whereas $n=2$ (red dashed with point) enhances the sensitivity significantly. The sensitivity with $\epsilon_{e\tau}$ decreases in the range $\delta_{CP}\sim[-\pi:-0.6\pi],~[0.4\pi:\pi]$ and increases for $\delta_{CP}\sim[-0.6\pi:0.4\pi]$ relative to SM. Therefore, the combined effect with $\epsilon_{e\tau} + \Gamma~(n = 2)$ enhances sensitivity compared to the combined effect with $\epsilon_{e\tau} + \Gamma~(n = 0)$.

In Appendix~\ref{appendix:Pmue-delcp-eps_etau}, the $\theta_{23}$ band in $P_{\mu e}$ versus $\delta_{CP}$ is plotted in fig.~\ref{fig_app:Pmue-delcp-eps_etau} to explain the reduction in MH sensitivity for true NH case. In the presence of $\epsilon_{e\tau}$, the NH band overlaps with the IH band for $\epsilon_{e\tau}'$. This degeneracy compromises the MH sensitivity. 
While the increase in MH sensitivity in the presence of decoherence can be explained by the $\nu_e$-appearance bump. For $n = 2$, we check an increase in $\nu_e$-appearance probability at high energy ($\sim 11$ GeV), leading to increase in events statistics in NH. However, the $P_{\mu e}$ for IH gets effected marginally. This causes increase in MH sensitivity for $n = 2$.

The CP-violating phase in the neutrino sector ($\delta_{CP}$) is yet to be measured precisely. 
CPV manifests itself as the difference in the probabilities of the appearance of neutrinos and antineutrinos. NSIs introduce additional non-standard phases ($\delta_{\alpha\beta}$), which mimic the effects of the intrinsic CPV phase ($\delta_{CP}$)~\cite{Liao:2016orc,Winter:2008eg}. Furthermore, decoherence also introduces a fake CP~\cite{Carpio:2019mass,Bera:2024hhr}. This causes degeneracies and reduces the sensitivity to CPV.

We define CPV sensitivity as
\begin{equation}\label{eq:cpv}
    \begin{aligned}
        \Delta\chi^2_{0,\pm\pi} &= \chi^2_{true}(\delta_{CP}^{true}, \epsilon^{true}_{e\tau}, \Gamma^{true}) - \chi^2_{test}(\delta_{CP}=0,\pm\pi, \epsilon^{test}_{e\tau}, \Gamma^{test})~,\\
    \Delta \chi_{CPV}^2 &= min [\Delta\chi_0^2, \Delta\chi_{\pm\pi}^2]~,
    \end{aligned}
\end{equation}
and marginalize over $\theta_{23}$, $\Delta m^2_{31}$, $\Gamma$, $\epsilon_{e\tau}$. We compute $\Delta\chi^2_{CPV}$ using true NH (IH) and plot as function of true $\delta_{CP}$ in the right upper (lower) panel of fig.~\ref{fig:cpv-mh-sensitivity-eps_etau}. The legends for different curves are as mentioned in the left plot. In top left plot we notice a small reduction of CPV sensitivity from standard sensitivity in the $\delta_{CP} \in [-\pi:0]$ for all scenarios. On the other hand in the $\delta_{CP} \in [0:\pi]$, the CPV sensitivity decreases w.r.t SM in presence of decoherence and increases in presence of $\epsilon_{e\tau}$. As a result, due to the combined effect the CPV sensitivity increases w.r.t SM for $\delta_{CP} \in [0:\pi]$.

The CPV sensitivity for IH is shown in the bottom right. We observe sensitivity decreases for all the cases in reference to the SM, except the sensitivity for $\epsilon_{e\tau}$ in the range $\delta_{CP}[0:\pi]$.

\FloatBarrier

\subsection{Impact of $\epsilon_{e\mu}$ and $\Gamma$ at DUNE far detector}
In this subsection, we present the effect of non-zero $\epsilon_{e\mu}$ and $\Gamma$ on neutrino propagation. Firstly, we discuss the modification of appearance probabilities and events rate. Using these modified probabilities, we study the MH and CPV sensitivities at DUNE.

\begin{figure*}[htb!]
\includegraphics[width=0.485\linewidth]{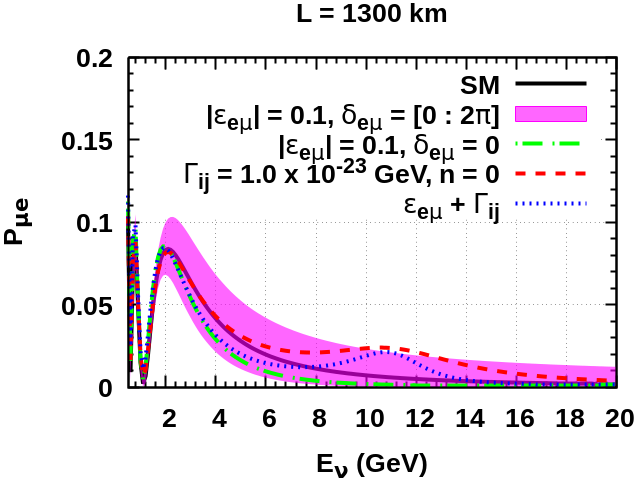}
\includegraphics[width=0.485\linewidth]{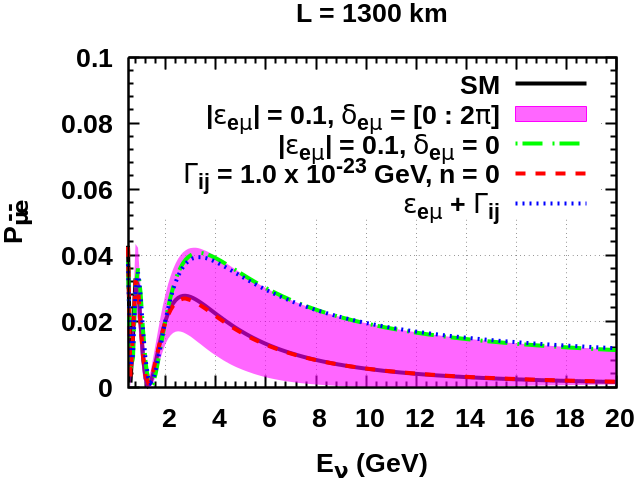}
\includegraphics[width=0.485\linewidth]{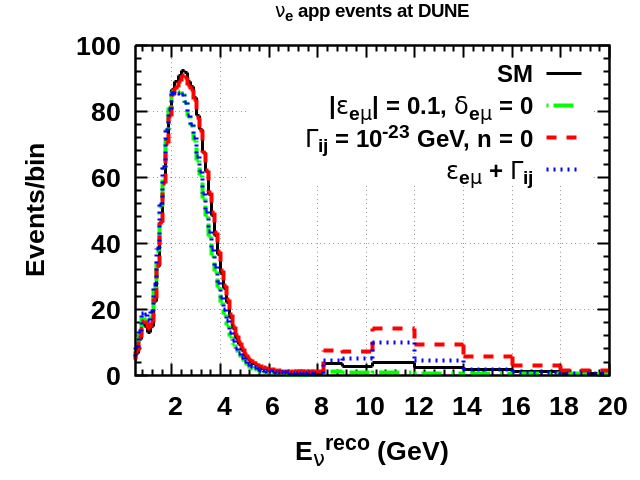}
\includegraphics[width=0.485\linewidth]{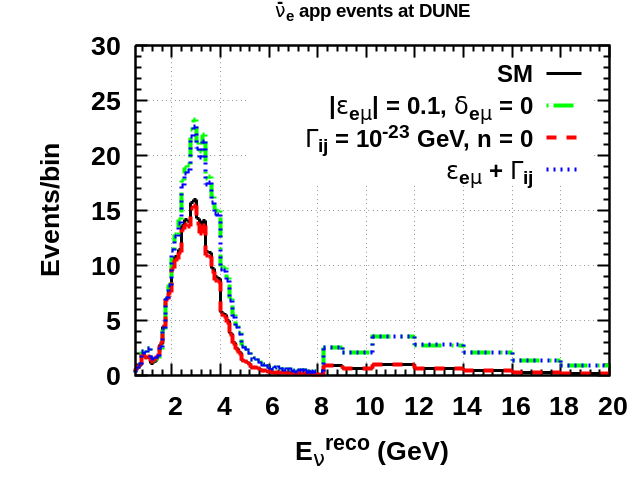}
\caption{Appearance probability (upper row) and appearance events spectrum (lower row) in the presence of $\epsilon_{e\mu}$ and $\Gamma$. Left and right panels correspond to neutrinos and antineutrinos respectively.}
\label{fig:prob-event-app-eps_emu}       
\end{figure*}
\subsubsection{Oscillation probabilities and events rate}
\noindent{\bf Appearance probability:} In the top row of fig.~\ref{fig:prob-event-app-eps_emu}, we display probabilities of $\nu_e$-appearance channel (left) and $\bar{\nu}_e$-appearance channel (right) considering NH. In each plot we show the following curves: black solid curve is for standard (SM) three flavor interaction; magenta band is for $|\epsilon_{e\mu}| = 0.1, \delta_{e\mu} = [0:2\pi]$; green dot-dashed line is for $|\epsilon_{e\mu}| = 0.1, \delta_{e\mu} = 0$; red dashed curve is for $\Gamma = 10^{-23}$ GeV with $n = 0$ and blue dotted curve is for combined effect of $\epsilon_{e\mu}$ and $\Gamma(n=0)$. 
The green curve shows that the probability with $\epsilon_{e\mu} = 0.1$ is lesser than the Standard Model (SM) case for $E_\nu > 3$ GeV. However, the observed magenta band indicates that other combinations of $|\epsilon_{e\mu}|$ and $\delta_{e\mu}$ result in $P_{\mu e}$ values exceeding those predicted by the SM. On the other hand, decoherence (red dashed curve) has minimal impact on $P_{\mu e}$ at low energies but produces a noticeable bump around 11 GeV. When considering the combined effects of both NSI and decoherence, the probability represented by the blue dotted curve is closer to the the green curve at lower energies and to the red curve at higher energies.

In the top right panel of fig.~\ref{fig:prob-event-app-eps_emu}, the red curve shows that decoherence does not affect the $\bar{\nu}_\mu \rightarrow \bar{\nu}_e$ channel, but the effect of $\epsilon_{e\mu}$ (green) is significant throughout the analysis energy window. As a result, the blue dotted curve representing the combined effect of NSI and decoherence is overlapping with green curve, thus prominently showcasing only the effect of NSI. 

\noindent{\bf Appearance events:} We present the appearance events in the bottom row of fig.~\ref{fig:prob-event-app-eps_emu}. The left and right plots correspond to the $\nu_e$ and $\bar{\nu}_e$ appearance signal events respectively. Four different curves in each plot represent the events/bin corresponding to the standard three flavor oscillations (black solid), NSI with $|\epsilon_{e\mu}| = 0.1$,$\delta_{e\mu} = 0$ (green dot-dashed), decoherence with $\Gamma = 10^{-23}$ GeV with $n=0$ (red dashed) and the combined effect (blue dotted). The trend followed by these events rate can be explained from the probability plots shown in top row. In table~\ref{table:event} we provide the number of events for the aforementioned cases. 
We have verified that the $\nu_\mu$ and $\bar{\nu}_\mu$ disappearance probabilities are not affected in the presence of $\epsilon_{e \mu}$, and the results are similar to fig~\ref{fig:prob-event-disapp-eps_etau}. 
\subsubsection{Mass hierarchy and CP-violation sensitivities}
\begin{figure*}
    \centering
    \includegraphics[width=0.485\linewidth]{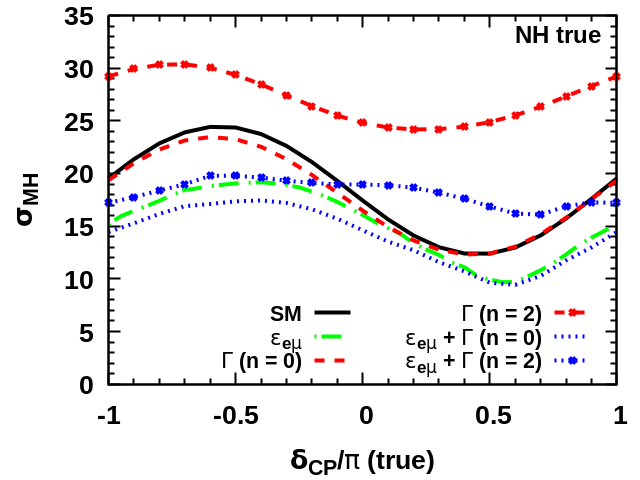}
    \includegraphics[width=0.485\linewidth]{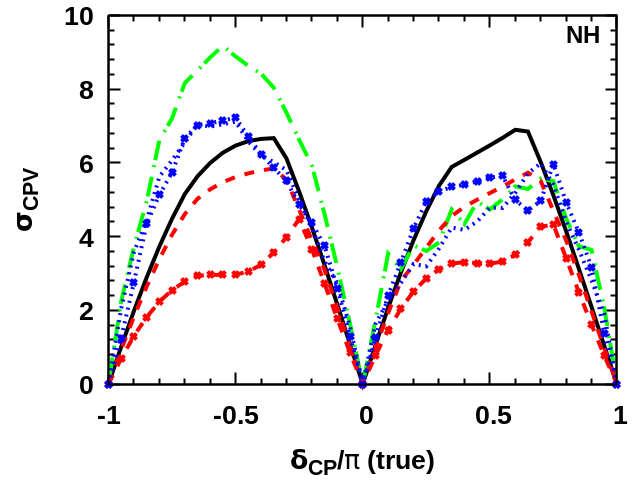}
    \includegraphics[width=0.485\linewidth]{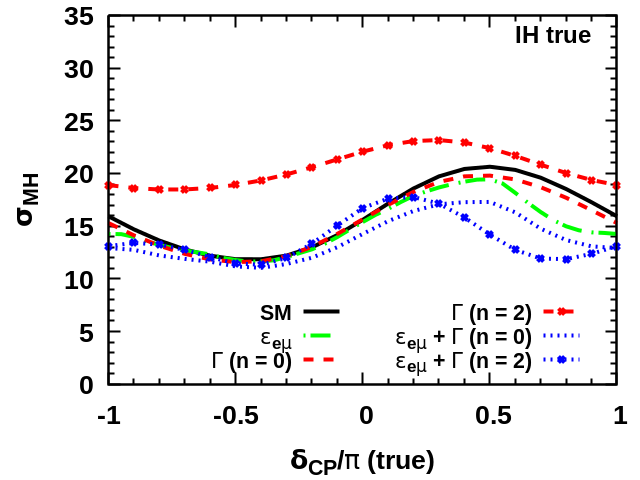}
    \includegraphics[width=0.485\linewidth]{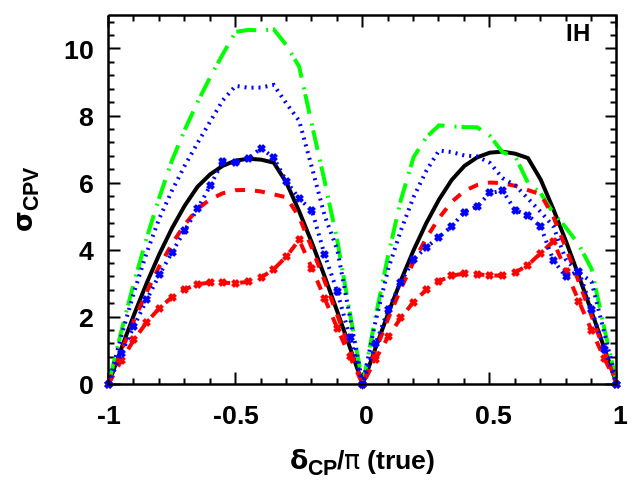}
    \caption{MH and CPV sensitivity as a function of $\delta_{CP}$. Upper and lower panel is for NH and IH. We consider true value of $\epsilon_{e\mu} = 0.1$, $\Gamma_0 = 10^{-23}$ GeV.}
    \label{fig:cpv-mh-sensitivity-eps_emu}
\end{figure*}

We now study the sensitivity to MH and CPV at DUNE in the presence of $\epsilon_{e\mu}$ and $\Gamma$.  
To obtain the sensitivities, 
we replace $\epsilon_{e\mu}$ with $\epsilon_{e\tau}$ in eq.~(\ref{eq:MH}) and (\ref{eq:cpv}) respectively for MH and CPV. In fig.~\ref{fig:cpv-mh-sensitivity-eps_emu} we illustrate MH and CPV sensitivity in left and right panel respectively. Upper and lower panels represent the results obtained assuming true hierarchy to be NH and IH respectively.

In the upper left plot of fig.~\ref{fig:cpv-mh-sensitivity-eps_emu} we illustrate the MH sensitivity considering NH as true hierarchy. In each plot we show sensitivity curves corresponding to SM (black), $\epsilon_{e\mu}$ (green), $\Gamma$ with $n = 0$ (red) and their combined effect (blue). The red and blue line points represent the $\Gamma$ with $n = 2$ and combined effect $\epsilon_{e\mu} + \Gamma(n = 2)$. From the green curve, assuming non-zero $\epsilon_{e\mu}$ in the data we see that the MH sensitivity reduces with respect to SM. When we assume decoherence in the data, we observe that the significance has marginally changed for the case of $\Gamma (n = 0)$ as seen from the red dashed line whereas it has hugely increased for $\Gamma (n = 2)$ as can be seen from the red line point.  
Once we consider the combined effect (blue dotted) the resultant sensitivity drops from SM throughout the true $\delta_{CP}\in [-\pi:\pi]$ for $\epsilon_{e\mu} + \Gamma(n = 0)$. At the same time, for $n = 2$ the sensitivity reduces for $\delta_{CP}\in[-\pi:0]$ and increases for the $\delta_{CP}\in[0:\pi]$ in comparison with SM as can be seen from blue line point.

In the bottom left plot, we show the MH sensitivity while taking IH as true hierarchy. The legends of different curves are same as in the upper plot. We can see from the red line point that there is an increase in the sensitivity w.r.t SM for individual effect of $\Gamma$ with power-law index $n = 2$. However, for rest of the cases the sensitivity doesn't deviate from the SM case (black curve) in the lower half plane (LHP) of $\delta_{CP}$. While in the upper half plane (UHP) of $\delta_{CP}$ the maximum deviation from SM occurs when we assume the combined effect of $\epsilon_{e\mu}+\Gamma(n = 2)$.

We illustrate the CPV sensitivity for NH in the upper right plot of fig.~\ref{fig:cpv-mh-sensitivity-eps_emu}. The legends of different curves are shown in the left plot. We notice that in the presence of $\epsilon_{e\mu}$ (green curve), CPV sensitivity is more than the SM case in the LHP and lesser in the UHP. While assuming $\Gamma$ with $n = 0, 2$ lowers the CPV sensitivity for the entire range of $\delta_{CP}$ w.r.t SM. The variation due to the combined effect is not significant relative to SM. The IH sensitivity presented in bottom right plot of fig.~\ref{fig:cpv-mh-sensitivity-eps_emu}, is higher than SM case for $\epsilon_{e\mu}$ and $\epsilon_{e\mu}+\Gamma(n=0)$ in the LHP as can be seen from the green and blue dotted lines. On the other hand, the raise in the sensitivity is not so significant in the UHP. Considering decoherence (red curves), the sensitivity decreases as the power-law indices increase. For the combined effect of $\epsilon_{e\mu}~{\text{and}}~\Gamma (n=2)$, sensitivity is almost similar to the SM case in the LHP and lesser in the UHP.

\FloatBarrier
\subsection{Differentiate NSI and decoherence}
The sensitivity of DUNE to distinguish between NSI and decoherence is discussed in the following section. To illustrate this, the statistical separation between NSI and decoherence is estimated by assuming one to be true and the other to be the test hypothesis. In this context, two distinct scenarios are considered. In the first scenario, NSI is assumed to be true, while decoherence is treated as the test hypothesis. In the second scenario, decoherence is assumed to be true, with NSI as the test hypothesis. The mass hierarchy is assumed to be normal in both simulated data and the test hypothesis. 

In the case of NSI, we have assumed only one non-zero off diagonal element at a time while differentiating with the decoherence phenomenon. That is, in subsection~\ref{sub:distin-etau-gamma} we have considered only non-zero $\epsilon_{e\tau}$ in the simulated data and tested with decoherence hypothesis and in subsection~\ref{sub:distin-emu-gamma} we explored non-zero $\epsilon_{e\mu}$ versus decoherence. Additionally, we have obtained the results assuming the 
the power-law dependency ($n = 0, 1, 2$) of the $\Gamma$ parameter.

    \begin{figure}
        \centering
        \includegraphics[width=0.485\linewidth]{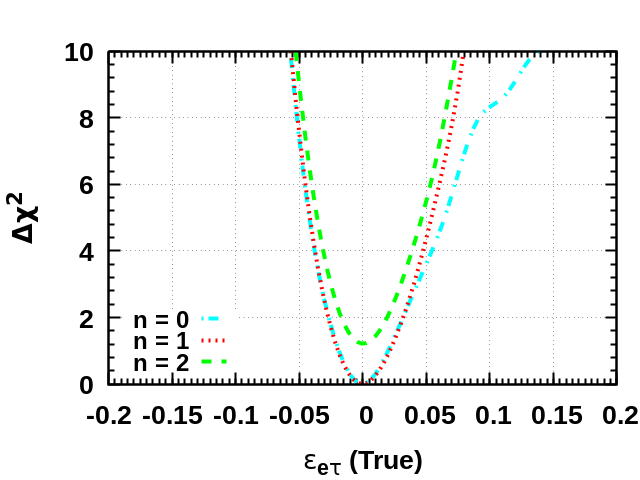}
        \includegraphics[width=0.485\linewidth]{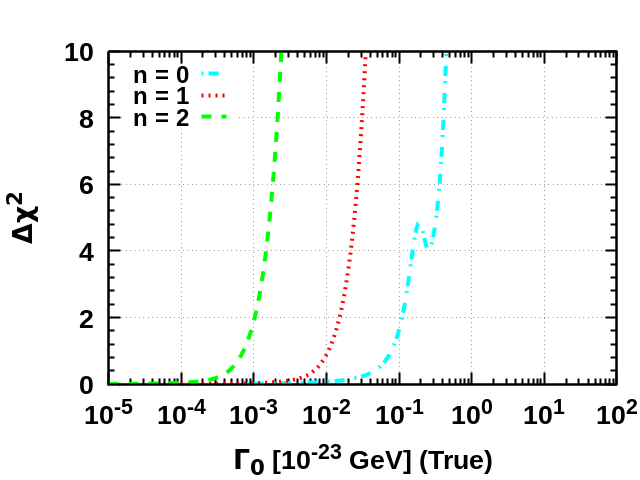}
        \caption{In the left (right) plot $\epsilon_{e\tau}$ (decoherence) is in the $true$ and decoherence ($\epsilon_{e\tau}$) is in the $test$.}
        \label{fig:differentiate-eps_etau-gamma}
    \end{figure}
    
\subsubsection{Distinguishing $\epsilon_{e\tau}$ and $\Gamma$:}\label{sub:distin-etau-gamma}

In this section, we generate the simulated data assuming the non-zero $\epsilon_{e\tau}$ in the range [-0.2, 0.2]. Fig.~\ref{fig:differentiate-eps_etau-gamma}, shows the significance with which DUNE can differentiate between NSI and decoherence for a certain true value of $\epsilon_{e\tau}$ obtained using eq.~(\ref{eq:eps_true-deco_test}).  
\begin{equation}\label{eq:eps_true-deco_test}
    \Delta \chi^2 = \chi^2_{true} (\epsilon_{e\tau} \neq 0, \Gamma = 0) - \chi^2_{test} (\epsilon_{e\tau} = 0, \Gamma \neq 0)~,
\end{equation}
Here we marginalize over $\theta_{23}$, $\delta_{CP}$ and $\Gamma$ in the test.
 The cyan, red, green lines corresponding to the different energy power-law dependencies $n = 0, 1, 2$ respectively on $\Gamma$ in the test statistics. We observe that for $n = 1, 2$, DUNE can differentiate $\epsilon_{e\tau}$ and $\Gamma$ with $3\sigma$ CL beyond this true range $-0.05<\epsilon_{e\tau}<0.07$. In the case of $n = 0$, for $\epsilon_{e\tau}<-0.05$ and $\epsilon_{e\tau}>0.13$ the two new physics can be distinguished with more than $3\sigma$ significance.

In the right panel of fig.~\ref{fig:differentiate-eps_etau-gamma} we show the second scenario where decoherence is assumed in true statistics and non-zero $\epsilon_{e\tau} \in [-0.19:0.14]$ in test statistics. To obtain the significance we have defined $\Delta \chi^2$ as 
\begin{equation}\label{eq:deco_true-eps_test}
    \Delta \chi^2 = \chi^2_{true} (\epsilon_{e\tau} = 0, \Gamma \neq 0) - \chi^2_{test} (\epsilon_{e\tau} \neq 0, \Gamma = 0)~,
\end{equation}
Here, we have marginalized over $\theta_{23}$, $\delta_{CP}$ and $\epsilon_{e\tau}$. The different power-law dependencies $n = 0, 1, 2$ are mentioned in the legends. 
For each of these curves, the $3\sigma$ value of $\Gamma_0$ beyond which the two NP scenarios are distinguished can be obtained by marking a horizontal line at $\Delta \chi^2 = 9$. In the case of  $n=0,1,2$ the $\Gamma_0$ values are $4.3 \times 10^{-24}$ GeV, $3.27 \times 10^{-24}$ GeV, $2.2 \times 10^{-26}$ GeV respectively, beyond which we can differentiate between $\epsilon_{e\tau}$ and $\Gamma$ with more than $3\sigma$ CL.

\subsubsection{Distinguishing $\epsilon_{e\mu}$ and $\Gamma$:}\label{sub:distin-emu-gamma}

 \begin{figure}[htb!]
        \centering
        \includegraphics[width=0.485\linewidth]{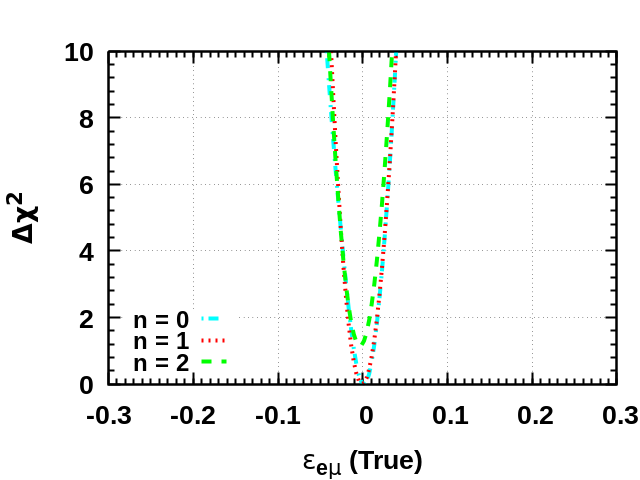}
        \includegraphics[width=0.485\linewidth]{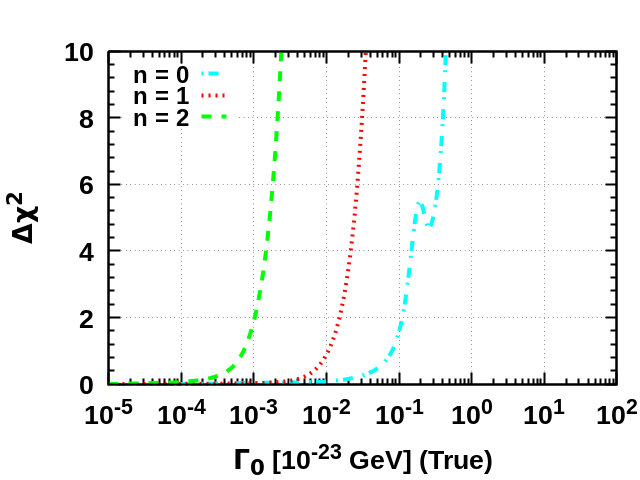}
        \caption{In the left (right) plot $\epsilon_{e\mu}$ (decoherence) is in the $true$ and decoherence ($\epsilon_{e\mu}$) is in the $test$.}
        \label{fig:differentiate-eps_emu-gamma}
    \end{figure}

In fig.~\ref{fig:differentiate-eps_emu-gamma} we differentiate between $\epsilon_{e\mu}$ and $\Gamma$. We consider $\Gamma$ depends on energy taking the power-law indices $n = 0, 1$ and $2$ as shown by the cyan, red and green curves, respectively. To compute the $\Delta\chi^2$ we proceed with the eq.~(\ref{eq:eps_true-deco_test}) and eq.~(\ref{eq:deco_true-eps_test}), where we replace $\epsilon_{e\tau}$ with $\epsilon_{e\mu}$.

In the left plot we illustrate $\Delta\chi^2$ as a function of true $\epsilon_{e\mu}$ taking non-zero $\Gamma$ in the test hypothesis. We observe for $-0.035 < \epsilon_{e\mu} < 0.035$ DUNE cannot differentiate between $\epsilon_{e\mu}$ and decoherence with $3\sigma$ CL. However, beyond $-0.035 < \epsilon_{e\mu} < 0.035$ leads to the $\Delta\chi^2 > 9$, indicating DUNE is sensitive to differentiate between $\epsilon_{e\mu}$ and $\Gamma$ with more than $3\sigma$ CL.

In the right plot of fig.~\ref{fig:differentiate-eps_emu-gamma} we distinguish $\epsilon_{e\mu}$ and $\Gamma$ assuming second scenario where we consider decoherence in the data and $\epsilon_{e\mu}$ in the test. We observe the similar trend as the right plot of fig.~\ref{fig:differentiate-eps_etau-gamma}. In the case of  $n=0,1,2$ the $\Gamma_0$ values are $4.3 \times 10^{-24}$ GeV, $3.27 \times 10^{-25}$ GeV, $2.2 \times 10^{-26}$ GeV respectively, beyond which we can differentiate between $\epsilon_{e\tau}$ and $\Gamma$ with more than $3\sigma$ CL. 


\section{Conclusions}\label{sec:conclusion}

In this work, we have analyzed the simultaneous effects of NSI and the environmental decoherence phenomenon on the neutrino propagation for the first time. We have investigated the sensitivity of the DUNE experiment to MH and CPV in the presence of the combined effect of both NP scenarios. For a comprehensive analysis, we have studied the MH and CPV sensitivity corresponding to the standard three flavor scenario and the individual impact of NSI and decoherence as well. In addition, we have demonstrated the sensitivity of DUNE to distinguish between NSI and the decoherence phenomenon. From the theoretical perspective, we have derived a general probability expression (eq.~\ref{eq:Pab}) that includes both NP parameters. We have shown the phenomenology of the dependency of energy on the decoherence parameters. Some noteworthy findings are summarized below.

\begin{itemize}
    \item We observe that around the peak beam energy of DUNE experiment i.e. 2.8 GeV, variation in $\epsilon_{e\tau}$ causes a notable change in the $P_{\mu e}$ value. Additionally, non-zero decoherence causes a peak at comparatively higher energy $\sim~11$ GeV.
    Therefore, the combined scenarios affect the probability at comparatively lower energy as well as at higher energy. On the contrary, in the case of $P_{\bar{\mu}\bar{e}}$ a non-zero decoherence doesn't affect the probability for the entire energy range. However, the combined analysis of $5(\nu)+5(\bar{\nu})$ years simulated data suggests a significant modification to the MH sensitivity when compared to the SM case (left panel of fig~\ref{fig:cpv-mh-sensitivity-eps_etau}). Similarly, in the case of CPV sensitivity (right panel of fig~\ref{fig:cpv-mh-sensitivity-eps_etau}), we see that the combined non-zero NP parameters compromise the sensitivity of DUNE around $\delta_{CP} = \pm \pi/2$.

    \item In the presence of $\epsilon_{e\mu}$ we notice that the $P_{\mu e}$ is affected throughout energy window. We observe similar trend in case of $P_{\bar{\mu}\bar{e}}$. When a non-zero $\epsilon_{e\mu}$ and $\Gamma$ are assumed in the simulated data we see that the combined parameters have lowered the hierarchy sensitivity in the case of $n = 0$ (left of panel of fig~\ref{fig:cpv-mh-sensitivity-eps_emu}).

    \item We have verified $\epsilon_{e\tau}$ and $\epsilon_{e\mu}$ do not affect the disappearance channels. Additionally, when non-zero decoherence is assumed, there is only a minimal change in the probability at relatively higher energy. Hence, the combined effect of two NP parameters is seen to be negligible.

    \item Assuming $\epsilon_{e\tau}$ in the simulated data and $\Gamma$ in the test hypothesis we observe that for $n = 1, 2$, DUNE can differentiate $\epsilon_{e\tau}$ and $\Gamma$ with $3\sigma$ CL beyond this true range $-0.05<\epsilon_{e\tau}<0.07$. In the case of $\Gamma$ in the true and $\epsilon_{e\tau}$ in the test hypothesis we see for $n=0,1,2$ the $\Gamma_0$ values are $4.3 \times 10^{-24}$ GeV, $3.27 \times 10^{-25}$ GeV, $2.2 \times 10^{-26}$ GeV respectively, beyond which we can differentiate between $\epsilon_{e\tau}$ and $\Gamma$ with more than $3\sigma$ CL. Similarly, we observe beyond $-0.035 < \epsilon_{e\mu} < 0.035$ in true and $\Gamma$ in the test leads to the $\Delta\chi^2 > 9$. While we assume $\Gamma$ in the true and $\epsilon_{e\mu}$ in the test hypothesis we see for $n=0,1,2$ the $\Gamma_0$ values are $4.3 \times 10^{-24}$ GeV, $3.27 \times 10^{-25}$ GeV, $2.2 \times 10^{-26}$ GeV respectively, beyond which we can differentiate between $\epsilon_{e\tau}$ and $\Gamma$ with more than $3\sigma$ CL.

\end{itemize}


\acknowledgments

\noindent CB thanks to Jochim Kopp for a fruitful discussion at \href{https://www.hri.res.in/~sangam/sangam24/index.html}{SANGAM @ HRI 2024}: Instructional Workshop in Particle Physics, Harish-Chandra Research Institute, Prayagraj, India. We thank to \href{https://physics.iith.ac.in/ppc2024/}{PPC 2024}: 17th International Conference on Interconnections between Particle Physics and Cosmology, for valuable comments from the participants while presenting the poster on this work. One of the authors (K. N. D.) thanks the Department of Science and Technology (DST)-SERB International Research Experience (SIRE) program for the financial support Grant No. SIR/2022/000518. We acknowledge high performance computing (HPC) facilities in \'Ecole Centrale School of Engineering - Mahindra University.
\FloatBarrier
\appendix
\section{Modified mixing matrix and mass-squared difference in presence of NSI}\label{app_mixing_mass_square}

In this appendix, we obtain the modified mixing matrix $\tilde{U}$ and mass square differences $\Delta \tilde{m}^2_{jk}$ using perturbation theory up to the first order of the small parameters $s_{13},~ \frac{\Delta m^2_{21}}{\Delta m^2_{31}}$, $\epsilon_{e\mu}$ and $\epsilon_{e\tau}$. 
The effective Hamiltonian including the non-standard interactions in eq.~(2.2) can be expanded as 
\begin{equation}\label{eqapp:H_eff}
    H_{eff} = \frac{\Delta m_{31}^2}{2E}\left(O_{23}U_\delta O_{13}U_\delta^\dagger O_{12} diag(0,\eta,1) O_{12}^\dagger U_\delta O_{13}^\dagger U_\delta^\dagger O_{23}^\dagger + \hat{A}~diag(1,0,0) + \hat{A}\epsilon \right)~,
\end{equation}
where $O_{12},O_{13},O_{23}$ are the rotational matrices, $U_\delta = diag(1,1,e^{i\delta})$, $\eta = \frac{\Delta m_{21}^2}{\Delta m_{31}^2}$, $\hat{A} = \frac{A}{\Delta m_{31}^2}$ and $\epsilon$ is a $3 \times 3$ matrix that contains only $\epsilon_{e\mu}$ and $\epsilon_{e\tau}$, rest all $\epsilon_{\alpha\beta}$ are set to zero. More rigorous expressions including all NSI parameters $\epsilon_{\alpha\beta} \neq 0$ are obtained in ref.~\cite{Meloni:2009ia}. 

\noindent The effective hamiltonian in eq.~\ref{eqapp:H_eff} can be rewritten in terms of matrix M as 
\begin{equation}
    H_{eff} = \frac{\Delta m_{31}^2}{2E}\left[O_{23} U_\delta M U_\delta^\dagger O_{23}^\dagger\right]~,
\end{equation}
where
\begin{equation}
    M = O_{13} U_\delta^\dagger O_{12} diag(0,\eta,1) O_{12}^\dagger U_\delta O_{13}^\dagger + U_\delta^\dagger O_{23}^\dagger diag(\hat{A},0,0) O_{23} U_\delta + \hat{A} U_\delta^\dagger O_{23}^\dagger \epsilon O_{23} U_\delta
\end{equation}

\begin{equation}
    M = M^{(0)} + M^{(1)} + ...~,
\end{equation}
where $M^{(i)}$ contains all terms of $i$th order in $\eta,~s_{13}$, $\epsilon_{e\mu}$ and $\epsilon_{e\tau}$.

\noindent Therefore,
\begin{equation}
    M^{(0)} = diag(\hat{A},0,1)
\end{equation}
and
\begin{equation}
    M^{(1)} = \begin{pmatrix}
        \eta s_{12}^2 & X & Y \\
        X^* & \eta c_{12}^2 & 0 \\
        Y^* & 0 & 0 
    \end{pmatrix}~,
\end{equation}
where $X = \frac{\Delta m^2_{21}}{\Delta m^2_{31}} s_{12}c_{12} + \frac{A}{\Delta m^2_{31}}\left(\epsilon_{e\mu}c_{23} - \epsilon_{e\tau}s_{23}\right)$ and $Y = s_{13} + \frac{A}{\Delta m^2_{31}} e^{i\delta_{CP}}(\epsilon_{e\mu}s_{23} + \epsilon_{e\tau}c_{23})$.

\noindent Eigenvalues and eigenvectors at the zeroth order: as $M^{(0)}$ is diagonal so
\begin{equation}\label{eqapp:lambda0}
\lambda_i^{(0)} = M_{ii}^{(0)},~ \hat{v}_i^{(0)} = e_i~.
\end{equation}

\noindent Eigenvalues and eigenvectors at the first order:
\begin{equation}\label{eqapp:lambda1}
    \lambda_i^{(1)} = M_{ii}^{(1)},~ \hat{v}_i^{(1)} = \sum_{j \neq i} \frac{M^{(1)}_{ji}}{\lambda_i^{(0)} - \lambda_j^{(0)}} \hat{v}_j^{(0)}
\end{equation}

\subsection{Modified mixing matrix}

Now, we obtain the modified mixing matrix $\tilde{U}$ using 
\begin{equation}
    \tilde{U} = O_{23} U_\delta \left(\hat{v}_1, \hat{v}_2, \hat{v}_3\right),
\end{equation}
where $\hat{v}_i = \hat{v}^{(0)}_i + \hat{v}^{(1)}_i$

\begin{equation}\label{eq:modified-PMNS}
    \tilde{U} = \begin{pmatrix}
        1 & \frac{X}{\hat{A}} & \frac{Y}{1-\hat{A}}e^{-i\delta_{CP}} \\
        c_{23}\frac{X^*}{\hat{A}} + s_{23} \frac{Y^*}{\hat{A} - 1}e^{i\delta_{CP}} & c_{23} & s_{23} \\
        -s_{23}\frac{X^*}{\hat{A}} + c_{23} \frac{Y^*}{\hat{A} - 1}e^{i\delta_{CP}} & -s_{23} & c_{23}
    \end{pmatrix}~.
\end{equation}

\subsection{Modified mass-squared differences}
Using the eigen values obtained in eq.~(\ref{eqapp:lambda0}) and eq.~(\ref{eqapp:lambda1}) we obtain the modified mass-squared differences below : 
\begin{equation}
    \tilde{m}_i^2 \approx \Delta m^2_{31} \left(\lambda_i^{(0)} + \lambda_i^{(1)}\right)
\end{equation}

\begin{equation}
    \begin{aligned}
        \Delta \tilde{m}^2_{21} &= \Delta m^2_{21}\cos(2\theta_{12}) - A \\
        \Delta \tilde{m}^2_{31} &= \Delta m^2_{31} - A - \Delta m^2_{21}\sin^2{\theta_{12}}
    \end{aligned}
    \label{eq:Delta_m_tilda}
\end{equation}
Here one can note that the $\Delta \tilde{m}^2_{21}$ and $\Delta \tilde{m}^2_{31}$ do not depend on the non-zero NSI parameters $\epsilon_{e\mu}$ and $\epsilon_{e\tau}$ that we have considered. 

\begin{figure*}[!htb]
    \centering
    \includegraphics[width=0.6\linewidth]{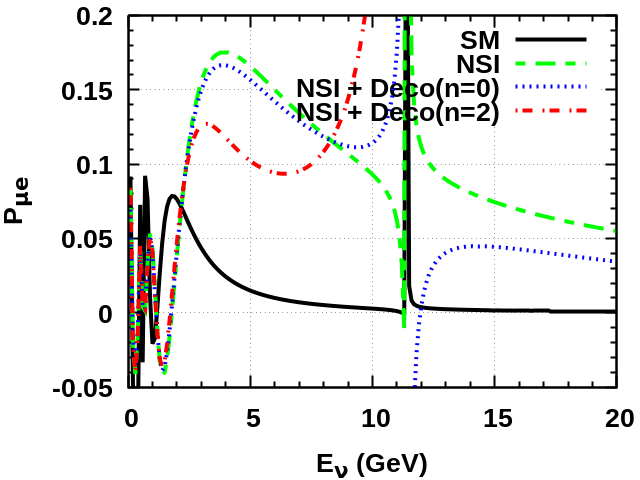}
    \caption{$P_{\mu e}$ versus $E_\nu$ using perturbative expression.}
    \label{fig:Pmue-perturbatio}
\end{figure*}

Substituting the modified parameters obtained via the first order perturbation (eqs.~(\ref{eq:modified-PMNS}), (\ref{eq:Delta_m_tilda})) in eq.~(2.9) we can obtain the appearance probability $P_{\mu e}$. We illustrate the $P_{\mu e}$ versus neutrino energy in fig.~\ref{fig:Pmue-perturbatio}. We observe that the probability encounters a singularity at resonance energy $\sim 11$ GeV (agrees with ref~\cite{Meloni:2009ia}). But this has been omitted in our numerical analysis, where we have obtained the exact probability expressions.


\section{Explanation of mass hierarchy sensitivity}\label{appendix:Pmue-delcp-eps_etau}

\begin{figure*}[!htb]
    \centering
    \includegraphics[width=0.458\linewidth]{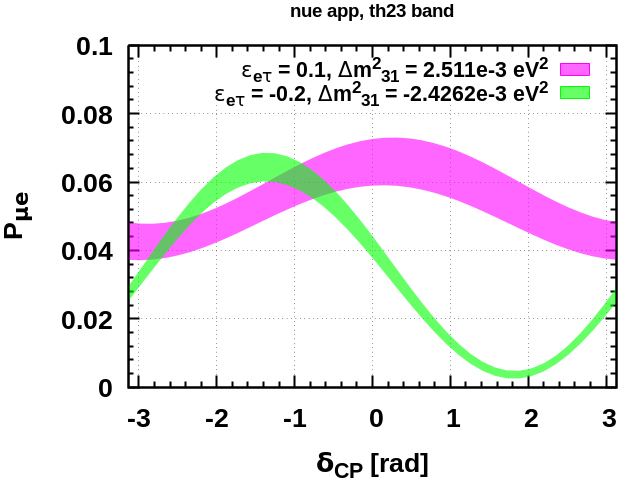}
    \includegraphics[width=0.458\linewidth]{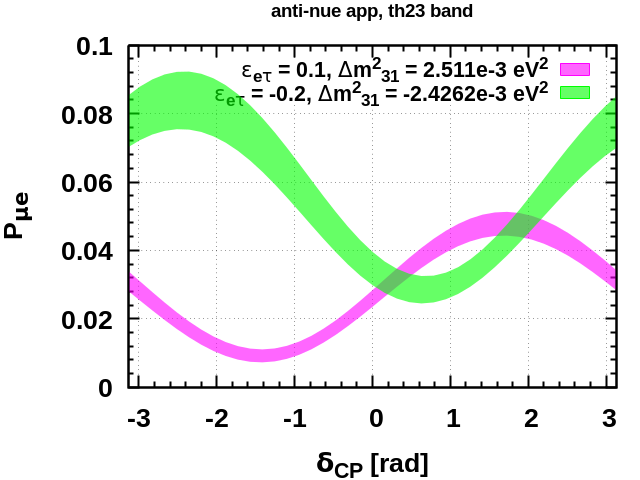}
    \caption{$\theta_{23}$ band in $P_{\mu e}$ - $\delta_{CP}$ plane.}
    \label{fig_app:Pmue-delcp-eps_etau}
\end{figure*}
To illustrate the decrease in the MH sensitivity in the presence of $\epsilon_{e\tau}$ we plot appearance probability $P_{\mu e}$ Vs $E_{\nu}$ in fig.~\ref{fig_app:Pmue-delcp-eps_etau}. The bands correspond to $3 \sigma$ range (Table-I) of $\theta_{23}$. In the left and right panels we show probabilities corresponding to $\nu_e$-appearance and $\bar{\nu}_e$-appearance channel. We represent true and test band with magenta and green colors respectively. We fix $\Delta m^2_{31}$ and $\epsilon_{e\tau}$ at $\Delta\chi^2_{min}$ to plot the probability band. 



\bibliographystyle{elsarticle-num}
\bibliography{bibliography.bib}
 
\end{document}